%% file: root.tex
\documentclass[letterpaper, 10 pt, conference]{ieeeconf}  

\IEEEoverridecommandlockouts                              

\usepackage{myStyle}
\usepackage{setspace}
\usepackage{eso-pic}
\usepackage{pgfornament}
\input{shortNames.tex}

\title{\LARGE \bf
Frequency Domain Identification of Multirate Systems: \\
 A Lifted Local Polynomial Modeling Approach
}

\author{Max van Haren$^{1}$, Lennart Blanken$^{1,2}$ and Tom Oomen$^{1,3}$
	\thanks{This work is part of the research programme VIDI with project number 15698, which is (partly) financed by the Netherlands Organisation for Scientific Research (NWO). In addition, this research has received funding from the ECSEL Joint Undertaking under grant agreement 101007311 (IMOCO4.E). The Joint Undertaking receives support from the European Union Horizon 2020 research and innovation programme.}
	\thanks{$^{1}$The authors are with the Control Systems Technology Section, Department of Mechanical Engineering, Eindhoven University of Technology, Eindhoven, The Netherlands.
		{\tt\small m.j.v.haren@tue.nl}}%
	\thanks{$^{2}$Lennart Blanken is with Sioux Technologies, Eindhoven, The Netherlands.}
	\thanks{$^{3}$Tom Oomen is with the Delft Center for Systems and Control, Delft University of Technology, Delft, The Netherlands.}}

\AddToShipoutPictureBG*{%
	\AtPageUpperLeft{%
		\setlength\unitlength{1in}%
		\hspace*{\dimexpr0.5\paperwidth\relax}
		\makebox(0,-0.75)[c]{\begin{tabular}{c c}
				Max van Haren, Frequency Domain Identification of Multirate Systems: A Lifted Local Polynomial Modeling Approach, \\
				To appear in {\em IEEE 61st Conference on Decision and Control}, Cancun, Mexico, 2022, uploaded to arXiv \today
		\end{tabular}}
}}

\begin{document}

\maketitle
\thispagestyle{empty}
\pagestyle{empty}

\begin{abstract}
Frequency-domain representations of multirate systems are essential for controller design and performance evaluation of multirate systems and sampled-data control. The aim of this paper is to develop a time-efficient closed-loop identification approach for multirate systems in the frequency-domain. The developed method utilizes local polynomial modeling for lifted representations of LPTV systems, which enables direct identification of closed-loop multirate systems in a single identification experiment. Unlike LTI identification techniques, the developed method does not suffer from bias due to ignored LPTV dynamics. The developed approach is demonstrated on a multirate example, resulting in accurate and fast identification in the frequency domain.
 
\end{abstract}
\input{Introduction}
\input{Preliminaries}
\input{ProblemDefinition}
\input{method}
\input{methodCL}
\input{Example}
\input{Conclusion}
\input{Appendix}
\def\url#1{}
\bibliographystyle{IEEEtran}
\bibliography{../../../library}

\end{document}

%% file: shortNames.tex
\providecommand{\uh}{\ensuremath{u_h}\xspace}
\providecommand{\rh}{\ensuremath{r_h}\xspace}
\providecommand{\Rh}{\ensuremath{R_h}\xspace}

\providecommand{\yh}{\ensuremath{y_h}\xspace}
\providecommand{\Yh}{\ensuremath{Y_h}\xspace}
\providecommand{\signalLength}{\ensuremath{N}\xspace}

\providecommand{\dt}{\ensuremath{n}\xspace}
\providecommand{\bin}{\ensuremath{k}\xspace}

\renewcommand{\uh}{\ensuremath{u_h}\xspace}
\renewcommand{\rh}{\ensuremath{r_h}\xspace}
\renewcommand{\Rh}{\ensuremath{R_h}\xspace}

\renewcommand{\Yh}{\ensuremath{Y_h}\xspace}

\renewcommand{\yh}{\ensuremath{y_h}\xspace}
\renewcommand{\signalLength}{\ensuremath{N}\xspace}

\renewcommand{\dt}{\ensuremath{n}\xspace}
\renewcommand{\bin}{\ensuremath{k}\xspace}

\providecommand{\ph}{\ensuremath{P_h}\xspace}
\providecommand{\kl}{\ensuremath{K_l}\xspace}
\providecommand{\izoh}{\ensuremath{I_{ZOH}}\xspace}
\providecommand{\hu}{\ensuremath{\mathcal{H}_u}\xspace}
\providecommand{\sd}{\ensuremath{\mathcal{S}_d}\xspace}

\providecommand{\nin}{\ensuremath{n_u}\xspace}
\providecommand{\nout}{\ensuremath{n_y}\xspace}

\renewcommand{\ph}{\ensuremath{P_h}\xspace}
\renewcommand{\kl}{\ensuremath{K_l}\xspace}
\renewcommand{\izoh}{\ensuremath{I_{ZOH}}\xspace}
\renewcommand{\hu}{\ensuremath{\mathcal{H}_u}\xspace}
\renewcommand{\sd}{\ensuremath{\mathcal{S}_d}\xspace}

\renewcommand{\nin}{\ensuremath{n_u}\xspace}
\renewcommand{\nout}{\ensuremath{n_y}\xspace}

\providecommand{\hh}{\ensuremath{h_h}\xspace}
\providecommand{\hl}{\ensuremath{h_l}\xspace}
\providecommand{\expjwL}{\ensuremath{e^{j\omega\hl}}\xspace}
\providecommand{\expjwH}{\ensuremath{e^{j\omega\hh}}\xspace}

\providecommand{\expjwkH}{\ensuremath{e^{j\omega_\fline\hh}}\xspace}
\providecommand{\lowfs}{\ensuremath{\omega_{s,l}}\xspace}
\providecommand{\highfs}{\ensuremath{\omega_{s,h}}\xspace}

\providecommand{\expjwk}{\ensuremath{e^{j\omega_\fline}}\xspace}
\providecommand{\expjw}{\ensuremath{e^{j\omega}}\xspace}
\providecommand{\expjwkr}{\ensuremath{e^{j\omega_{\fline+\window}}}\xspace}

\renewcommand{\hh}{\ensuremath{h_h}\xspace}
\renewcommand{\hl}{\ensuremath{h_l}\xspace}
\renewcommand{\expjwL}{\ensuremath{e^{j\omega\hl}}\xspace}
\renewcommand{\expjwH}{\ensuremath{e^{j\omega\hh}}\xspace}
\renewcommand{\lowfs}{\ensuremath{\omega_{s,l}}\xspace}
\renewcommand{\highfs}{\ensuremath{\omega_{s,h}}\xspace}

\renewcommand{\expjwkH}{\ensuremath{e^{j\omega_\fline\hh}}\xspace}

\renewcommand{\expjwk}{\ensuremath{e^{j\omega_\fline}}\xspace}
\renewcommand{\expjw}{\ensuremath{e^{j\omega}}\xspace}
\renewcommand{\expjwkr}{\ensuremath{e^{j\omega_{\fline+\window}}}\xspace}

\providecommand{\lift}{\ensuremath{\mathcal{L}}\xspace}
\providecommand{\fac}{\ensuremath{F}\xspace}

\renewcommand{\lift}{\ensuremath{\mathcal{L}}\xspace}
\renewcommand{\fac}{\ensuremath{F}\xspace}

\providecommand{\fline}{\ensuremath{k}\xspace}
\providecommand{\window}{\ensuremath{r}\xspace}
\providecommand{\windowSize}{\ensuremath{n}\xspace}

\renewcommand{\fline}{\ensuremath{k}\xspace}
\renewcommand{\window}{\ensuremath{r}\xspace}
\renewcommand{\windowSize}{\ensuremath{n}\xspace}

\DeclareDocumentCommand{\theoremRef}{m}{Theorem~\ref{#1}}
\DeclareDocumentCommand{\figRef}{m}{Fig.~\ref{#1}}
\DeclareDocumentCommand{\secRef}{m}{Section~\ref{#1}}
\DeclareDocumentCommand{\lemmaRef}{m}{Lemma~\ref{#1}}
\DeclareDocumentCommand{\defRef}{m}{Definition~\ref{#1}}


%% file: Introduction.tex
\section{Introduction}
Multirate sampling is becoming more relevant due to increasing performance requirements and system complexity, for example in sampled-data control \cite{Chen1995}, or networked control systems \cite{Hespanha2007}, where multiple interacting loops are sampled at different rates. The multirate sampling results in Linear Periodically Time Varying (LPTV) dynamics \cite{Bittanti2009}. Control design techniques for multirate systems are hampered due to the lack of Linear Time Invariant (LTI) properties. \par

Frequency-domain representations are an important role for performance evaluation and control design of LTI systems, e.g., as is used in manual loop-shaping techniques \cite{Schmidt2020} and parametric identification \cite{Pintelon1994}. Frequency-domain representation is typically done by Frequency Response Functions (FRFs). The major benefit of FRFs is that they are non-parametric, i.e., no prior knowledge about the system is necessary to identify FRFs. Additionally, FRFs are directly determined from input-output data, identified fast, and inexpensive \cite{Pintelon2012}. Finally, FRFs enable the direct evaluation of stability, performance and robustness margins \cite{Skogestad2007}. \par

Frequency-response methods for performance evaluation and control design of LTI systems cannot be directly applied to LPTV systems, since, in contrast to LTI systems, the frequency separation principle does not hold \cite{Oomen2007,VanZundert2019}. Consider for example multirate systems, where a single input frequency influences multiple output frequencies, and a single output frequency is influenced by multiple input frequencies. In \cite{Uyanik2019,Yin2020} equivalent LTI representations are used to identify LPTV systems, however they do not consider frequency-domain representations or closed-loop systems. Alternative frequency-domain representations for sampled-data control are developed in \cite{Araki1996,Yamamoto1996}, and include the Performance Frequency Gain (PFG) \cite{Lindgarde1997}, which can readily be used for performance evaluation. The equivalent description of the PFG for multirate control is developed in \cite{Oomen2007}. However, identification of the PFG requires a model of the high-rate plant, which is not trivial to identify, or an identification experiments for each individual input frequency is necessary, which is time-consuming. Hence, no time-efficient methods to identify frequency-domain representations of closed-loop multirate systems are present. \par

Although multirate systems and controller implementations are broadly used, systematic frequency-domain identification techniques for these implementations are not yet available. The aim of this paper is to develop a fast, accurate, and inexpensive frequency-response identification approach for closed-loop multirate systems. The key idea to overcome the lack of the frequency-separation principle is to use time-invariant representations of LPTV systems \cite{Kranc1957,Bittanti2009}, that are capable of representing LPTV systems in the frequency domain. The time-invariant representations transform an LPTV system into a larger dimensional LTI system by lifting over time or frequency. Additionally, Local Polynomial Modeling (LPM) \cite{Pintelon2012}, that exploits local smoothness of transfer functions, is employed to identify the time-invariant representations of LPTV systems in a single identification experiment. Furthermore, a high-rate plant and the PFG for multirate systems are determined using the identified time-invariant representations through an inverse-lifting procedure. The contributions include:
\begin{itemize}
	\item[C1] Single experiment identification of time-invariant representations for LPTV systems by lifting input-output data and applying LPM (\secRef{sec:method}).
	\item[C2] The use of these time-invariant representations to identify high-rate plants and the PFG for multirate systems operating in closed-loop, e.g., for the use in sampled-data control systems (\secRef{sec:methodCL}).
	\item[C3] The developed framework is validated on a closed-loop multirate example (\secRef{sec:Example}).
\end{itemize}

%% file: Preliminaries.tex
\section{Preliminaries}
\label{sec:preliminaries}
In this section, the notation used in this paper and a description for LPTV systems are presented.
\subsection{Notation}
Considered systems have \nin inputs and \nout outputs. All systems are discrete, and the frequency response is denoted with $G(\expjw)$. Sampled discrete-time signals are denoted as $\nu[\dt]$, with discrete time $\dt\in \{ 0,1,\ldots,\signalLength-1\}$, and \signalLength the total amount of samples. The operator $q$ is the forward shift operator, i.e., $q\nu[\dt]=\nu[\dt+1]$. Discrete-time Fourier transforms of signals are given by $V(\expjw)$, with $\omega\in[0,2\pi)$. The discrete-time Fourier transform based on finite amount of samples $\omega_\bin=\frac{2\pi}{\signalLength}\bin, \; \forall \bin \in \{0,1,\ldots,\signalLength-1\}$ is given by the Discrete Fourier Transform (DFT) \cite[Section~2.2.2]{Pintelon2012}
\begin{equation*}
	\begin{split}
		V(\bin)=\frac{1}{\sqrt{\signalLength}}\sum_{\dt=0}^{\signalLength-1}\nu[\dt]e^{-j2\pi\dt\bin/\signalLength},
	\end{split}
\end{equation*}
with \bin the \bin'th frequency bin. The $(i,j)$'th element of matrix $A$ is denoted as $A_{[i,j]}$ and the conjugate transpose as $A^{H}$.

\subsection{Description of LPTV Systems}
The input-output behavior of LPTV system $G_{LPTV}(q,\dt)$, evaluated at time \dt, is given by the impulse response \cite{Bittanti2009}
\begin{equation}
	\label{eq:LPTVIO}
	\begin{split}
		y[\dt] &= \sum_{i=0}^{\infty}M_i[\dt]q^{-i}\,u[\dt],\\
		 &= G_{LPTV}(q,\dt)\,u[\dt]
	\end{split}
\end{equation}
where the matrix coefficients $M_i[\dt] \in \mathbb{R}^{\nout\times\nin}$ are $\fac$-periodic functions, i.e., $M_i[\dt]=M_i[\dt+\fac]$.

%% file: ProblemDefinition.tex
\section{Problem Formulation}
\label{sec:Problem}
In this section, the problem considered is presented. First, the problem setup is given. Second, the problem associated to the identification of FRFs for multirate systems is shown. Finally, the problem addressed in this paper is defined.
\subsection{Problem Setup}
\label{sec:problemsetup}
\begin{figure}[tb]
	\centering
	\includegraphics[width=\linewidth]{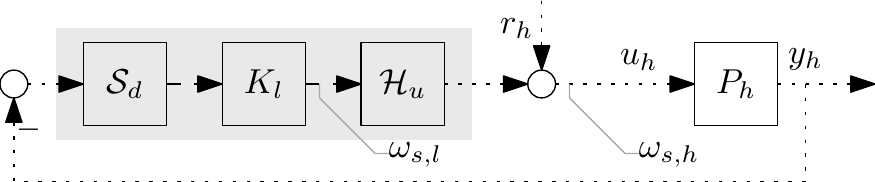}
	\caption{Multirate feedback structure for system identification considered in this paper, with downsampler \sd, low-rate controller \kl, a multirate zero-order-hold reconstructor $\hu$ and high-rate plant \ph.}
	\label{fig:MRFeedbackSystem}
\end{figure}

In this paper, the multirate feedback control structure in \figRef{fig:MRFeedbackSystem} is considered, where the plant \ph is sampled on a high-rate $\highfs=\frac{2\pi}{\hh}$, operating in closed-loop with a stabilizing controller \kl on a low-rate $\lowfs=\frac{2\pi}{\hl}$. The sampling times of the low- and high-rate are related by $\hl=\fac\hh$, with $\fac\in\mathbb{N}$. Note that more than two sampling rates are possible, however, they should have a common divisor. 
The downsampler and multirate zero-order-hold reconstructor are defined in \cite{Vaidyanathan1993} and \cite{Oomen2007}, respectively.
\subsection{
Frequency-Domain Representation of Multirate Systems
}
\label{sec:MRFRF}
Consider the multirate closed-loop structure as in \figRef{fig:MRFeedbackSystem}, where the output \yh can be described as
\begin{equation}
	\yh = J_{LPTV}\,\rh,
\end{equation}
which is given in the frequency domain as
\begin{equation}
	\label{eq:yh}
	\resizebox{0.91\hsize}{!}{%
		$\begin{aligned}
			&\Yh(\expjwH) = \ph(\expjwH)\Rh(\expjwH)-\ph(\expjwH)I_{ZOH}(\expjwH) \\
			&\cdot Q_d(\expjwL)\frac{1}{\fac}\sum_{f=0}^{\fac-1}\ph\left( e^{j\hh\left( \omega-\left( f/\fac\right) \highfs\right)}\right)\Rh(e^{j\hh\left(\omega-\left( f/\fac\right) \highfs \right) }),  
		\end{aligned}$}
\end{equation}
with
\begin{equation}
	\label{eq:Qd}
		\resizebox{0.91\hsize}{!}{%
		$\begin{aligned}
			Q_d(\expjwL) &= \left(I+\kl\left(\expjwL\right) P_l\left(\expjwL\right)\right)^{-1} \kl\left(\expjwL\right),\\
			\izoh(\expjwH) &=  \sum_{f=0}^{\fac-1}e^{-j\omega \hh f}.
		\end{aligned}$}
\end{equation}
From \eqref{eq:yh} it is clear that the frequency separation principle does not hold, since \Yh, for a single frequency, is influenced by \fac frequencies from \Rh, since the down- and upsampler alias the feedback signal. In Example~\ref{example:output}, this is illustrated.
\begin{example}
	\label{example:output}
	Consider the multirate closed-loop in \figRef{fig:MRFeedbackSystem}, with $\fac=3$. The system is excited with $\Rh(e^{j\omega_0\hh})=0.5$, with $\omega_0=2\pi\cdot60$ rad/s. The output \Yh contains multiple frequencies, as seen in \figRef{fig:ExampleOutput}. 
	\begin{figure}[tb]
		\centering\includegraphics{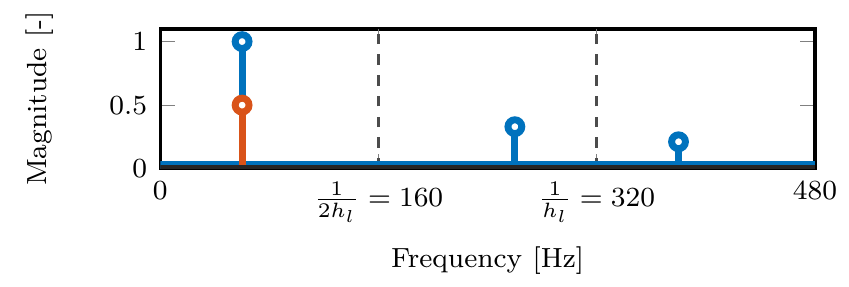}
		\caption{\Yh \li{blue}{solid}[1.4] and \Rh \li{red}{solid}[1.4] for the multirate system with $\fac=3$.}
		\label{fig:ExampleOutput}
	\end{figure}
	After applying transfer function estimate $\frac{\Yh}{\Rh}$, estimation errors occur around 260 and 380 Hz, since there is zero input but non-zero output.
\end{example}

For multirate systems, alternative frequency-domain representations exist, including the PFG, seen in \defRef{definition:regPFG}.
\begin{definition}[PFG \cite{Oomen2007,Lindgarde1997}]
	\label{definition:regPFG}
	The PFG $\mathcal{P}$, with performance and exogenous variables $\zeta\in\mathbb{R}^{n_\zeta}$ and $\eta\in\mathbb{R}^{n_\eta}$, is
	\begin{equation}
		\begin{aligned}
			\mathcal{P}\left(\expjwkH\right)=\sup_{\eta \neq 0} \frac{\left\|\zeta\right\|_{\mathcal{P}}}{\left\|\eta\right\|_{\mathcal{P}}},
		\end{aligned}
	\end{equation}
	given that $\eta\in\mathcal{W}_P$ and
	\begin{align}
		\mathcal{W}_P&=\left\{\eta\left[\dt \right],  \eta\left[\dt \right]=c \expjwkH,\|c\|_{2}<\infty\right\}, \\
		\|\cdot\|_{\mathcal{P}}&=\sqrt{\lim _{N \rightarrow \infty} \frac{1}{2 N+1} \sum_{\dt=-N}^{N}\|\cdot\|^{2}},
	\end{align}
	with $\|\cdot\|$ the Euclidean vector norm and meaning that $\eta$ can only contain a single frequency.	
\end{definition}
The PFG, in the case of sampled-data control, represents the full intersample behavior of a system.
\subsection{Problem Definition}
Identification techniques for multirate systems are hampered due to the lack of the frequency-separation principle. First, closed-loop identification of the high-rate plant \ph is not trivial, due to the lack of the frequency separation principle. Second, experimental identification of the PFG is time-consuming since the excitation signal can only contain a single frequency to circumvent the frequency-separation principle. Therefore, the aim of this paper is to develop an identification procedure that identifies frequency-domain representations of multirate systems in a single experiment.

%% file: method.tex
\section{LPM Identification of Time-Invariant Representations of LPTV systems}
\label{sec:method}
In this section, LPTV systems are identified in a single experiment, leading to contribution C1. First, time-invariant representations of LPTV systems from literature are presented, that transform LPTV systems into higher dimensional LTI systems. Second, the time-invariant representations are identified in a single experiment by lifting the input-output data of LPTV systems and exploiting LPM. 

\subsection{Time-Invariant Representations of LPTV Systems}
\label{sec:timeInvariantRep}
Time-invariant representations are used to represent LPTV in the frequency domain, where in this paper, the time-lifted and frequency-lifted reformulations are considered. First, the time-lifted reformulation gathers the values of a signal from one period into an augmented signal, i.e., the time-lifted signal \cite{Kranc1957,Ohnishi2021}, as given in given by \defRef{def:timeLiftSig}.
\begin{definition}[Time-lifted signals]
	\label{def:timeLiftSig}
	Given a signal $\nu_h[\dt]\in\mathbb{R}^{n_\nu}$, the time-lifted signal $\underline{\nu}[l]=\lift\nu_h\left[ \dt\right] \in \mathbb{R}^{\fac n_\nu}$ with lifting operator \lift, is given by
	\begin{equation}
		\label{eq:timeLiftSignal}
		\resizebox{0.99\hsize}{!}{%
			$\begin{aligned}
				\underline{\nu}\left[ l \right]  &=\begin{bmatrix}
					\nu_h^\top\left[ l\fac\right]  & \nu_h^\top\left[ l\fac+1\right] &\ldots&\nu_h^\top\left[ l\fac+\fac-1\right] 
				\end{bmatrix}^\top.
			\end{aligned}$}
	\end{equation}
\end{definition}
Using time-lifted signals, LPTV systems are transformed into LTI systems, as seen in \lemmaRef{lemma:timelift}. \par
\begin{lemma}[Time-lifted reformulation of LPTV systems]
	\label{lemma:timelift}
	The time-lifted representation of $G_{LPTV}$ from \eqref{eq:LPTVIO} is
	\begin{equation}
		\label{eq:underlineG}
		\underline{G}=\lift G_{LPTV} \lift^{-1}\in\mathcal{R}^{\fac\nout\times\fac\nin},
	\end{equation}
	having LTI input-output behavior
	\begin{equation}
		\underline{y} = \underline{G}\,\underline{u}.
	\end{equation}
	In the frequency domain, the elements of $\underline{G}$ are given by
	\begin{equation}
		\label{eq:GunderlineRep}
		\scalebox{0.86}{$\displaystyle	
			\underline{G}_{[p,q]}\left(\expjw\right)=\frac{e^{j\omega(p-q)\frac{1}{\fac}}}{\fac} \sum_{f=0}^{\fac-1} G_{LPTV}\left(e^{j\omega\frac{1}{\fac}} \phi^{f\frac{1}{\fac}}, p-1\right) \phi^{f(p-q)\frac{1}{\fac}},
			$}
	\end{equation}
	where $p,q \in \{1,2,\ldots,\fac\}$, $\phi=e^{j\frac{2 \pi}{F}}$. and $G_{LPTV}$ from \eqref{eq:LPTVIO}.
\end{lemma}
\begin{proof}
	For a proof, see, e.g., \cite[Section~8.3]{Chen1995} and \cite[Section~6.2.2]{Bittanti2009}.
\end{proof}
Second, lifting is also done directly in the frequency-domain \cite{Zhang1996}. Frequency-lifted signals are given by \defRef{def:freqLiftSig}
\begin{definition}[Frequency-lifted signals]
	\label{def:freqLiftSig}
	Given a discrete-time Fourier transform of a signal $V_h(\expjw)\in\mathbb{C}^{n_\nu}$, the frequency-lifted signal $\tilde{V}(\expjw)\in \mathbb{C}^{\fac n_\nu}$ is given by
	\begin{equation}
		\label{eq:freqLiftSignal}
		\scalebox{0.85}{$\displaystyle
			\tilde{V}(\expjw)=\begin{bmatrix}
			V_h^\top(\expjw) &
			V_h^\top(\expjw \phi) &
			\cdots &
			V_h^\top\left(\expjw \phi^{\fac-1}\right)
		\end{bmatrix}^\top.
	$}
	\end{equation}
\end{definition}
The elements of $\tilde{V}$ include aliased signals of the original signal $V_h$. Using frequency-lifted signals, LPTV systems are transformed into LTI systems, as seen in \lemmaRef{lemma:freqlift}.
\begin{lemma}[Frequency-lifted reformulation of LPTV systems]
	\label{lemma:freqlift}
	The frequency-lifted representation of LPTV system $G_{LPTV}$ from \eqref{eq:LPTVIO} is given by 
	\begin{equation}
		\label{eq:tildeG}
\tilde{G}(z)=M(z)\underline{G}(z^\fac)M^{-1}(z) \in\mathcal{R}^{\fac\nout\times\fac\nin},
	\end{equation}
	where $z=\expjw$, 
	\begin{equation}
		\footnotesize M(z)= \left[\begin{array}{cccc}
			I & z^{-1}I & \cdots & e^{\left( -\fac+1\right)\omega }I \\
			I & (z \phi)^{-1} & \cdots & (z \phi)^{-\fac+1} \\
			\vdots & \vdots & \cdots & \vdots \\
			I & \left(z \phi^{\fac-1}\right)^{-1} & \cdots & \left(z \phi^{\fac-1}\right)^{-\fac+1}
		\end{array}\right], \normalsize %
	\end{equation}
	and having LTI input-output behavior
	\begin{equation}
		\tilde{y} = \tilde{G}\,\tilde{u}.
	\end{equation}
\end{lemma}
\begin{proof}
	The proof follows from \lemmaRef{lemma:timelift} and the fact that $M$ is an LTI operator \cite[Section~6.4]{Bittanti2009}.
\end{proof}
\subsection{LPM for Time-Invariant Representations of LPTV Systems}
\label{sec:LPM}
Ideally, the FRFs of LPTV systems have to be identified in a single identification experiment. In this paper, the input-output data of LPTV systems is lifted, resulting in an equivalent multivariable LTI system. Consequently, LPM is utilized to identify the multivariable LTI system in a single experiment, which is possible since LPM exploits the local smoothness of transfer functions. \par
Define the time-lifted or frequency-lifted input in the frequency domain as $\underline{U}$, and the output as $\underline{Y}$, which, for open-loop lifted system $\underline{G}(\expjwk)\in\mathbb{C}^{\fac\nout\times\fac\nin}$, is equal to
\begin{equation}
	\label{eq:LPMoutput}
	\underline{Y}(\fline) = \underline{G}(\expjwk)\underline{U}(\fline) + \underline{V}(\fline) + \underline{T}(\expjwk),
\end{equation}
with noise contribution $\underline{V}(\fline)$ and transient contribution $\underline{T}(\expjwk)$, combining both the noise and system transient. A local window \window can be taken around it in the frequency domain, ranging from $\window \in \{-\windowSize,\ldots,\windowSize\}$, i.e.,
\begin{equation}
	\label{eq:windowedOutput1}
	\underline{Y}(\fline+\window) = \underline{G}(\expjwkr)\underline{U}(\fline+\window) + \underline{V}(\fline+\window) + \underline{T}(\expjwkr),
\end{equation}
where the lifted systems are approximated using polynomials $\underline{g}_s(\fline)$ and $\underline{t}_s(\fline)$ \cite{Pintelon2012}, i.e.,
\begin{equation}
	\begin{aligned}
		&\underline{G}\left(e^{j \omega_{k+r}}\right) \approx \underline{G}(\expjwk) + \textstyle\sum_{s=1}^{R}\underline{g}_s(\fline)r^s, \\
		&\underline{T}\left(e^{j \omega_{k+r}}\right) \approx \underline{T}(\expjwk) + \textstyle\sum_{s=1}^{R}\underline{t}_s(\fline)r^s,
	\end{aligned}
\end{equation}
where $R$ is typically chosen as 2. These approximations are now used in \eqref{eq:windowedOutput1}, resulting in
\begin{equation}
	\label{eq:windowedOutput}
		\underline{Y}(\fline+\window) \approx \theta K(\fline+\window) + \underline{V}(\fline+\window),
\end{equation}
with $\theta \in \mathbb{C}^{\fac\nout \times \left( R+1\right)\left(\fac\nin+1\right)}$ containing $\underline{G}(\expjwk)$, $\underline{T}(\expjwk)$, $\underline{g}_s(\bin)$ and $\underline{t}_s(\bin)$ and vector $K(\fline+\window)\in\mathbb{C}^{\left( R+1\right)\left(\fac\nin+1\right)}$ is
\begin{equation}
	\resizebox{0.87\hsize}{!}{%
	$\begin{aligned}
		K(\fline+\window) = \begin{bmatrix}
			K_1(\window)\otimes \underline{U}(\fline+\window) \\
			K_1(\window)
		\end{bmatrix} &&  \text{with } 	K_1 = \begin{bmatrix}
			1 \\ r \\ \vdots \\ r^R
		\end{bmatrix},
	\end{aligned}$}
\end{equation}
with $\otimes$ the Kronecker product. 
Collecting input-output data results in the least-squares optimum system estimate $\widehat{\underline{G}}$ as
\begin{equation}
	\widehat{\underline{G}}(\expjwk) = \left( \underline{Y}_nK_n^H\left(K_nK_n^H \right) ^{-1}\right) \begin{bmatrix}
		I_{\fac\nin} &
		0 &
		\cdots &
		0
	\end{bmatrix}^\top,
\end{equation}
with $X_n = \begin{bmatrix}X(\fline-\windowSize) & \ldots & X(\fline+\windowSize)\end{bmatrix}$. For more details the reader is referred to \cite[Section~7.2]{Pintelon2012}.
\begin{remark}
	\label{remark:LPM}
	Since DFT bins $\fline-\windowSize,\ldots,\fline+\windowSize$ are used for estimating $\widehat{\underline{G}}(\expjwk)$, it means that $\widehat{\underline{G}}(\expjwk)$ is correlated with $\widehat{\underline{G}}(\expjwkr)$ for $\window = -2\windowSize,\ldots,2\windowSize$. This explains why more parameters ($\fac\nout\times(R+1)(\fac\nin+1)$) are able to be identified than the number of measurements $\fac\nout$, if
	\begin{equation}
		\label{eq:LPMreq}
		2\windowSize+1-(R+1)(\fac\nin+1)>\fac\nout.
	\end{equation}
	 Hence, time-lifted or frequency-lifted systems can be identified in a single experiment.
\end{remark}

%% file: methodCL.tex
\section{Multirate Closed-Loop Identification of high-rate plant \ph and the PFG}
\label{sec:methodCL}
In this section, the method is developed that identifies high-rate plant \ph and the PFG, operating in multirate closed-loop as seen in \figRef{fig:MRFeedbackSystem}, leading to contribution C2. First, lifted representations of the high-rate plant \ph are calculated by exploiting the result of \secRef{sec:method}. Second, inverse lifting is applied to identify the high-rate plant \ph, that subsequently is used to compute the PFG. Third, it is shown that the bias error of the developed method does not contain the multirate effects illustrated in \eqref{eq:yh}, in contrast to LTI identification techniques. Fourth, the bias errors of the lifting techniques are compared, showing that the frequency-lifted reformulation is less sensitive to bias errors introduced by LPM. Finally, the entire developed procedure is presented.
\subsection{Identification of Time-Lifted and Frequency-Lifted \ph}
Time-lifted and frequency-lifted representations of high-rate plant \ph are identified using an indirect method. First, define the LPTV systems
\begin{align}
	\label{eq:JLPTV}
	\yh=J_{LPTV}\,\rh &= \left( \left(I+\ph \hu \kl \sd\right)^{-1} \ph\right) \,\rh, \\
	\label{eq:SLPTV}
	\uh=S_{LPTV}\,\rh &= \left(I+ \hu \kl \sd\ph\right)^{-1}\,\rh,
\end{align}
that have lifted representations $\underline{J}$, $\tilde{J}$, $\underline{S}$ and $\tilde{S}$, as seen in \lemmaRef{lemma:timelift} and \lemmaRef{lemma:freqlift}. Consecutively, lifted representations of the high-rate plant \ph are calculated using \theoremRef{theorem:liftP}. 
\begin{theorem}[Calculating $\underline{P}$ and $\tilde{P}$]
	\label{theorem:liftP}
	Given the lifted representations $\underline{J}$, $\underline{S}$, $\tilde{J}$ and $\tilde{S}$, the lifted plant is obtained using the indirect method \cite[Section~2.6.4]{Pintelon2012} as
	\begin{align}
		\label{eq:liftedP}
		\underline{P}=\underline{J}\, \underline{S}^{-1}, && \tilde{P}=\tilde{J}\, \tilde{S}^{-1}.
	\end{align}
\end{theorem}
\begin{proof}
	Use the push-through rule \cite[Section~3.2]{Skogestad2007} on $\underline{J}$ and substitute, with $\underline{S}$, into the right side of \eqref{eq:liftedP}, i.e.,
	\begin{equation}
			\underline{J} = \lift\ph\left(I+\hu \kl \sd\ph\right)^{-1}  \lift^{-1},
	\end{equation}
	 and substitute, together with $\underline{S}$, into the right side of  \eqref{eq:liftedP}, e.g., for time lifting this is equal to
	\begin{equation}
		\resizebox{0.88\hsize}{!}{%
		$\begin{aligned}
			\underline{J}\, \underline{S}^{-1} &= \lift\ph\left(I+\hu \kl \sd\ph\right)^{-1}  \lift^{-1} \lift\left(I+\hu \kl \sd \ph\right)  \lift^{-1} \\
			 &= \lift \ph\lift^{-1} = \underline{P},
		\end{aligned}$}
	\end{equation}
	and the frequency-lifted case follows directly from this.
\end{proof}
The lifted representations $\underline{P}$ and $\tilde{P}$, in combination with inverse lifting, result in the high-rate plant.
\subsection{Calculation of High-Rate Plant \ph and PFG}
The identified time-lifted and frequency-lifted representations of \ph are used to calculate the high-rate plant \ph and the PFG. First, the inverse time-lifting and frequency-lifting procedures for LTI systems, seen in respectively \lemmaRef{lemma:invTimeLift} and \lemmaRef{lemma:invFreqLift}, are used to estimate the high-rate plant \ph, given $\underline{P}$ and $\tilde{P}$ calculated with \theoremRef{theorem:liftP}.
\begin{lemma}[Inverse time-lifting for LTI systems]
	\label{lemma:invTimeLift}
 	The inverse time-lifting procedure for LTI systems is given by
	\begin{equation}
		\label{eq:inverseTimeLift}
		\ph(\expjwH) = \sum_{\sigma=0}^{\fac-1}P^{(\sigma)}\left(e^{j\omega\hl\fac}\right) e^{-\sigma j\omega\hh},
	\end{equation}
	where $P^{(\sigma)}$ is the $\sigma$'th element of the first column of $\underline{P}$.
\end{lemma}
\begin{proof}
	For a proof, see {\cite[Section~6.2.1]{Bittanti2009}}.
\end{proof}
\lemmaRef{lemma:invTimeLift} shows that, due to the lower Nyquist frequency of the time-lifted representation, a single frequency of $P^{(\sigma)}$ influences \fac frequencies of the high-rate plant \ph.
\begin{lemma}[Inverse frequency-lifting for LTI systems]
	\label{lemma:invFreqLift} 
	Inverse frequency-lifting for LTI systems is given by
		\begin{equation}
				\label{eq:inverseFreqLift}
				\ph(\expjwH) =\tilde{P}(\expjwH\phi^{-p})_{\left[ p+1,p+1\right] },
		\end{equation}
	where $p\in\{0,1,\ldots,\fac-1\}$ can be chosen.
\end{lemma}
\begin{proof}
	For a proof, see {\cite[Section~6.4.1]{Bittanti2009}}.
\end{proof}
In contrast to time lifting, \lemmaRef{lemma:invFreqLift} shows that a single frequency of the frequency-lifted representation $\tilde{P}$ influences a single frequency of the high-rate plant \ph. Finally, the PFG is calculated differently compared to \defRef{definition:regPFG}, by using the identified high-rate plant \ph as seen in \cite{Oomen2007}. 
\subsection{Bias Error for Identifying \ph}
The bias error when identifying \ph does not contain LPTV effects, in contrast to LTI identification techniques. However, LPM introduces an interpolation and leakage bias. The bias errors on the lifted plants $\underline{P}$ and $\tilde{P}$, shown in the Appendix, are directly used to calculate the bias on the high-rate plant \ph, as seen in \theoremRef{theorem:biasPhTime} and \theoremRef{theorem:biasPhFreq}.
\begin{theorem}[Bias error on \ph when using time-lifted reformulation]
	\label{theorem:biasPhTime}
	The bias error on estimation of \ph for time-lifted reformulations of LPTV systems is given by
	\begin{equation}
		\label{eq:biasPhTime}
		\resizebox{0.88\hsize}{!}{%
			$
			\begin{aligned}
				&\mathbb{E}\left( \widehat{\ph}\left( \expjwH\right) \right) -\ph\left( \expjwH\right)  =\sum_{\sigma=0}^{\fac-1}\bigg( \Big(\underline{J}_\Delta\left( \expjwL\right) \\
				&-\underline{P}\left( \expjwL\right) {\underline{S}}_\Delta\left( \expjwL\right)  \Big) \underline{S}^{-1}\left( \expjwL\right) \bigg)_{[\sigma+1,1]} e^{-\sigma j\omega\hh},
			\end{aligned}
		$}
	\end{equation}
with $X_\Delta$ defined in \eqref{eq:xdelta}, consisting of a polynomial interpolation and leakage error.
\end{theorem}
\begin{proof}
	Substitute the inverse lifting procedure, seen in \eqref{eq:inverseTimeLift}, into the left side of \eqref{eq:biasPhTime} and apply \lemmaRef{lemma:biasP}, i.e.,
	\begin{equation}
		\resizebox{0.88\hsize}{!}{%
			$\begin{aligned}
				&\mathbb{E}\left( \widehat{\ph}\left( \expjwH\right)\right) -\ph\left( \expjwH\right)= \\
				 &\sum_{\sigma=0}^{\fac-1}	\mathbb{E}\left(\hat{P}^{(\sigma)}\left(e^{j\omega\hl\fac}\right)\right)  e^{-\sigma j\omega\hh} -\sum_{\sigma=0}^{\fac-1}P^{(\sigma)}\left(e^{j\omega\hl\fac}\right) e^{-\sigma j\omega\hh}=\\
				&\sum_{\sigma=0}^{\fac-1}\left( \mathbb{E}\left(\widehat{\underline{P}}_{[\sigma+1,1]}\left(e^{j\omega\hl\fac}\right)\right) -{\underline{P}}_{[\sigma+1,1]}\left(e^{j\omega\hl\fac}\right)\right) e^{-\sigma j\omega\hh},
			\end{aligned}$}
	\end{equation}
that, in combination with \eqref{eq:biasLift}, results in \eqref{eq:biasPhTime}.
\end{proof}
The bias error on estimating the high-rate plant \ph using time lifting does not contain LPTV effects. Furthermore, a single frequency of the bias error of the time-lifted elements $\underline{P}_{[\sigma+1,1]}$ influences \fac frequencies of the high-rate plant \ph, due to the lowered Nyquist frequency of the time-lifted representation, and can also be concluded from \eqref{eq:inverseTimeLift}.
\begin{theorem}[Bias error on \ph when using frequency-lifted reformulation]
	\label{theorem:biasPhFreq}
	The bias error on estimation of \ph for frequency-lifted reformulations of LPTV systems is
	\begin{equation}
		\label{eq:biasPhFreq}
			\begin{split}
				&\mathbb{E}\left(\widehat{\ph}\left( \expjwH\right)\right) -\ph\left( \expjwH\right)= \bigg( \Big({\tilde{J}}_\Delta\left( \expjwH\right)\\
				&-\tilde{P}\left( \expjwH\right){\tilde{S}}_\Delta\left( \expjwH\right) \Big) \tilde{S}^{-1}\left( \expjwH\right)\bigg)_{[p+1,p+1]}
			\end{split}
	\end{equation}
	where $p\in\{0,1,\ldots,\fac-1\}$ can be freely chosen.
\end{theorem}
\begin{proof}
	In the left side of \eqref{eq:biasPhFreq}, substitute the estimated and true inverse lifting procedure from \eqref{eq:inverseFreqLift}, i.e.,
	\begin{equation}
			\begin{aligned}
				&\mathbb{E}\left(\widehat{\ph}\left( \expjwH\right)\right) -\ph\left( \expjwH\right)= \\
				 	&\mathbb{E}\left(\widehat{\tilde{P}}(\expjwH\phi^{-p})_{\left[ p+1,p+1\right] }\right) -\tilde{P}(\expjwH\phi^{-p})_{\left[ p+1,p+1\right] },
			\end{aligned}
	\end{equation}
which is equal to \eqref{eq:biasPhFreq} when applying \lemmaRef{lemma:biasP}.
\end{proof}
Similar to time lifting, the bias on identifying the high-rate plant \ph by using frequency lifting does not contain LPTV effects, but solely polynomial interpolation and leakage effects. In contrast to time lifting, the bias of a single frequency of the frequency-lifted reformulation influences only a single frequency of the high-rate plant \ph. This leads to the main result of this paper, a method that is not biased due to LPTV dynamics and does not repeat in the frequency domain.

\subsection{Procedure of Developed Method}
The developed procedure identifies high-rate model \ph and the PFG in a single experiment by applying LPM and time-invariant representations, resulting in no bias due to the LPTV dynamics, and is summarized in in Algorithm~\ref{alg:method}.
\begin{algorithm}[ht]
	\DontPrintSemicolon
	\KwIn{\fac, \highfs}
	\KwOut{$\widehat{\underline{P}}$, $\widehat{\tilde{P}}$, $\widehat{\ph}$, $\widehat{\mathcal{P}}$}
	Excite multirate feedback loop in \figRef{fig:MRFeedbackSystem} with suitably designed \rh, covering the entire frequency range \;
	Apply time and frequency lifting to signals \rh, \uh and \yh and take the DFT into $\underline{R}$, $\underline{U}$, $\underline{Y}$, $\tilde{R}$, $\tilde{U}$ and $\tilde{Y}$ using \defRef{def:timeLiftSig} and \defRef{def:freqLiftSig}.\;
	Perform LPM on lifted signals to identify lifted representations $\widehat{\underline{J}}$, $\widehat{\underline{S}}$, $\widehat{\tilde{J}}$ and $\widehat{\tilde{S}}$, see \secRef{sec:LPM}. \;
	Use \theoremRef{theorem:liftP} to estimate $\widehat{\underline{P}}$ and $\widehat{\tilde{P}}$.\;
	Apply inverse lifting procedure from \lemmaRef{lemma:invTimeLift} and \lemmaRef{lemma:invFreqLift} on $\widehat{\underline{P}}$ and $\widehat{\tilde{P}}$, resulting in $\widehat{\ph}$. \;
	Calculate $\widehat{\mathcal{P}}$ with $\widehat{\ph}$ as described in \cite{Oomen2007}.\;
	\caption{Developed method for identifying high-rate plant and the PFG of multirate system.}
	\label{alg:method}
\end{algorithm}


%% file: Example.tex
\section{Example}
\label{sec:Example}
In this section, the developed method is validated on an example, leading to contribution C3. First, the considered system is introduced. Second, the compared methods are presented. Finally, the results of the example are given.
\subsection{System Description}
\begin{figure}[tb]
	\centering
	\includegraphics{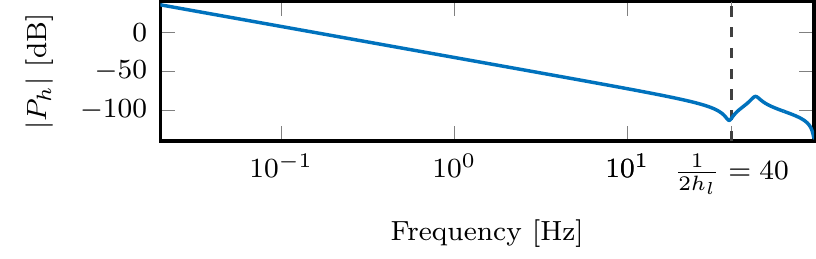}
\caption{Bode magnitude diagram of the example plant \ph.}
	\label{fig:examplePlantController}
\end{figure}

A mass-spring-damper system is used, where a bode magnitude is seen in \figRef{fig:examplePlantController}. The sampling frequencies are $\highfs=240\cdot 2\pi$ and $\lowfs=80\cdot2\pi$ rad/s, i.e., $\fac=3,$. Data is gathered during 6000 s, by choosing \rh as white noise. 
Note that \ph has a resonance frequency above the low Nyquist frequency. Furthermore, the polynomial degree $R$ in LPM has been chosen as 2, and the window length $n=8$, such that $2\windowSize+1-(R+1)(\fac\nin+1)>\fac\nout$, as seen in \eqref{eq:LPMreq}.
\subsection{Compared methods}
The estimation of the high-rate plant \ph and the PFG is compared for several methods, namely:
\begin{itemize}
	\item (ETFE) Using an Emperical Transfer Function Estimate, without a window, such that the frequency resolution is the same as for the other methods.
	\item (LPM) Using LPM directly on the high-rate data \rh, \uh and \yh, i.e. neglecting the LPTV dynamics.
	\item (Time lifted) Identifying time-lifted systems $\underline{J}$ and $\underline{S}$, which are used to calculate \ph and the PFG.
	\item (Frequency lifted) Identifying frequency-lifted systems $\tilde{J}$ and $\tilde{S}$, which are used to calculate \ph and the PFG.
\end{itemize}
\subsection{Results}
\label{sec:exampleResults}
\begin{figure}[tb]
	\centering\includegraphics{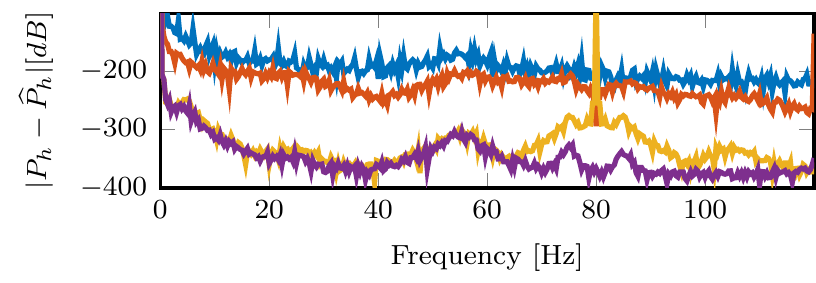}
	\caption{Modeling error $|\ph-\widehat{\ph}|$ for ETFE \li{blue}{solid}[1.5pt], LPM \li{red}{solid}[1.5pt], Time-lifted \li{yellow}{solid}[1.5pt] and frequency-lifted \li{purple}{solid}[1.5pt]. 
	}
	\label{fig:ModelError}
\end{figure}

\begin{figure}[tb]
	\centering\includegraphics{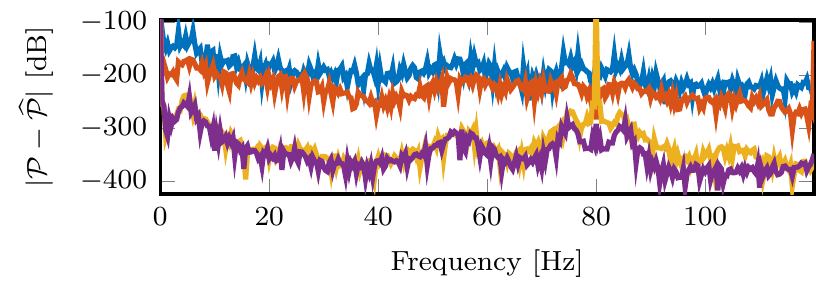}
	\caption{Error on estimation of the PFG for ETFE \li{blue}{solid}[1.5pt], LPM \li{red}{solid}[1.5pt], Time lifted \li{yellow}{solid}[1.5pt] and frequency lifted \li{purple}{solid}[1.5pt]. 
	}
	\label{fig:PFGError}
\end{figure}

The framework identifies high-rate plant \ph and the PFG for the multirate closed-loop. The achieved modeling error on \ph and the PFG using the compared methods can be seen in \figRef{fig:ModelError} and \figRef{fig:PFGError}. 
The following observations are made.
\begin{itemize}
	\item Time lifting outperforms ETFE and LPM in terms of the modeling error for most frequencies, but performs worse around 80 Hz. This can be explained with \eqref{eq:biasPhTime}, that shows bias errors of the lifted representations repeat in the frequency-domain for the high-rate plant, in this case the error for low frequencies is repeated.
	\item The modeling error of the developed method using frequency lifting outperforms the other methods, since it does not ignore LPTV effects and the bias introduced by LPM is not repeated in the frequency domain.
\end{itemize}

%% file: Conclusion.tex
\section{Conclusions}
\label{sec:Conclusion}
In this paper, a method is developed to identify the PFG for multirate systems operating in closed loop. Multivariable time-invariant representations of closed-loop transfer functions are identified in a single experiment by applying LPM. The original high-rate system is found by inverse lifting. Finally, the original system is used to calculate the PFG. A benchmark example shows that the estimation of both the original system and the PFG has improved significantly compared to neglecting LPTV dynamics of multirate systems. \par
Ongoing research is directed at exploiting the additional structure which is present in time- or frequency-lifted representations to improve the LPM estimate. Lastly, experimental validation of the framework is ongoing work.

%% file: Appendix.tex
\section*{Appendix - Bias Errors on Lifted Systems}
\label{sec:appendix}
The bias on lifted LPTV systems is shown in \lemmaRef{lemma:biasLift}.
\begin{lemma}[Bias error in lifted transfer functions]
	\label{lemma:biasLift}
	The bias errors that are introduced by LPM when identifying lifted transfer functions are given by
	\begin{equation}
		\begin{aligned}
			\mathbb{E}\left(\widehat{\underline{G}}  \right)-\underline{G} = \underline{G}_\Delta
		\end{aligned}
	\end{equation}
	where,
	\begin{equation}
		\label{eq:xdelta}
		\scalebox{0.85}{$\displaystyle	
			X_\Delta={X}^{(R+1)} \mathcal{O}_{\text{int} {X}}\left((\windowSize / \signalLength)^{(R+1)}\right) +\mathcal{O}_{\text{leak} {X}}\left((\windowSize / \signalLength)^{(R+2)}\right), \\
		$}
	\end{equation}
	and $X^{(R+1)}$ the $(R+1)$'th derivative of $X$ with respect to frequency, interpolation and leakage error $\mathcal{O}_{\text{int}X}$ and $\mathcal{O}_{\text{leak}X}$.
\end{lemma}
\begin{proof}
	For a proof, see \cite[Section~7.2.2]{Pintelon2012}.
\end{proof}
Second, the bias on $\underline{P}$ and $\tilde{P}$ is given in \lemmaRef{lemma:biasP}.
\begin{lemma}[Bias errors on lifted plants]
	\label{lemma:biasP}
	The bias error on estimating the time-lifted or frequency-lifted plant $\underline{P}$ or $\tilde{P}$ operating in closed-loop is approximately equal to
	\begin{equation}
		\label{eq:biasLift}
		\begin{aligned}
			\mathbb{E}\left( \widehat{\underline{P}}\right)-\underline{P}=&\left(\underline{J}_\Delta-\underline{P}\,{\underline{S}}_\Delta \right) \underline{S}^{-1}, \\
			\mathbb{E}\left( \widehat{\tilde{P}}\right)-\tilde{P}=&\left({\tilde{J}}_\Delta-\tilde{P}{\tilde{S}}_\Delta \right) \tilde{S}^{-1}.
		\end{aligned}
	\end{equation}
\end{lemma}
\begin{proof}
	For a proof, see {\cite[Appendix~7.F]{Pintelon2012}}.
\end{proof}